\documentclass{emulateapj}
\usepackage{natbib}
\usepackage{graphicx}
\usepackage{multirow}
\usepackage{longtable}

\defcitealias{Brown98}{B98}
\defcitealias{Brown00}{Brown et al. 2000}

\newcommand{\AGBm}{\mbox{AGB-manqu\'e}}
\newcommand{\pagb}{\mbox{P-AGB}}
\newcommand{\peagb}{\mbox{PE-AGB}}
\newcommand{\hphb}{\mbox{HP-HB}}
\newcommand{\hhb}{\mbox{EHB}}
\newcommand{\Msun}{\mbox{$M_{\odot}$}}
\newcommand{\Lsun}{\mbox{$L_{\odot}$}}
\newcommand{\Teff}{\mbox{$T_{\rm eff}$}}

\newcommand{\um}{\mbox{$\mu$m}}
\newcommand{\uvb}{UV-bright}
\newcommand{\sersic}{S\'ersic}

\shorttitle{Bright UV Stars in the Bulge of M31}
\shortauthors{Rosenfield et al.}

\begin{document}

\title{The Panchromatic Hubble Andromeda Treasury I: Bright UV Stars in the Bulge of M31\footnote{Based on observations made with the NASA/ESA Hubble Space Telescope, obtained from the Data Archive at the Space Telescope Science Institute, which is operated by the Association of Universities for Research in Astronomy, Inc., under NASA contract NAS 5-26555.}}

\author{
Philip Rosenfield\altaffilmark{1},
L. Clifton Johnson\altaffilmark{1},
L\'eo Girardi\altaffilmark{2}, 
Julianne J.\ Dalcanton\altaffilmark{1},
Alessandro Bressan\altaffilmark{16}, 
Dustin Lang\altaffilmark{3},
Benjamin F. Williams\altaffilmark{1},
Puragra Guhathakurta\altaffilmark{5},
Kirsten M. Howley\altaffilmark{15},
Tod R.\ Lauer\altaffilmark{9},
Eric F.\ Bell\altaffilmark{14},
Luciana Bianchi\altaffilmark{12}
Nelson Caldwell\altaffilmark{7},
Andrew Dolphin\altaffilmark{6}, 
Claire E. Dorman\altaffilmark{5}
Karoline M. Gilbert\altaffilmark{1,13},
Jason Kalirai\altaffilmark{11},
S{\o}ren S. Larsen\altaffilmark{17},
Knut A.G. Olsen\altaffilmark{9},
Hans-Walter Rix\altaffilmark{10}
Anil C. Seth\altaffilmark{4}, 
Evan D.\ Skillman\altaffilmark{8}
Daniel R. Weisz\altaffilmark{1}
}

\altaffiltext{1}{Department of Astronomy, University of Washington, Box 351580, Seattle, WA 98195, USA}
\altaffiltext{2}{Osservatorio Astronomico di Padova -- INAF, Vicolo dell'Osservatorio 5, I-35122 Padova, Italy}
\altaffiltext{3}{Department of Astrophysical Sciences, Princeton University, Princeton, NJ 08544}
\altaffiltext{4}{Department of Physics \& Astronomy, University of Utah, Salt Lake City, UT 84112}
\altaffiltext{5}{University of California Observatories/Lick Observatory, University of California, 1156 High St., Santa Cruz, CA 95064}
\altaffiltext{6}{Raytheon Company, 1151 East Hermans Road, Tucson, AZ 85756, USA}
\altaffiltext{7}{Harvard-Smithsonian Center for Astrophysics, 60 Garden Street Cambridge, MA 02138, USA}
\altaffiltext{8}{Minnesota Institute for Astrophysics, University of Minnesota, 116 Church Street SE, Minneapolis, MN 55455, USA}
\altaffiltext{9}{National Optical Astronomy Observatory, 950 North Cherry Avenue, Tucson, AZ 85719, USA}
\altaffiltext{10}{Max Planck Institute for Astronomy, Koenigstuhl 17, 69117 Heidelberg, Germany}
\altaffiltext{11}{Space Telescope Science Institute, 3700 San Martin Drive, Baltimore, MD, 21218, USA}
\altaffiltext{12}{Department of Physics and Astronomy,  Johns Hopkins University, Baltimore, MD 21218, USA}
\altaffiltext{13}{Hubble Fellow}
\altaffiltext{14}{Department of Astronomy, University of Michigan, 500 Church St., Ann Arbor, MI 48109, USA}
\altaffiltext{15}{Lawrence Livermore National Laboratory, 7000 East Avenue, Livermore, CA 94550, USA}
\altaffiltext{16}{SISSA, via Bonomea 265, 34136 Trieste, IT}
\altaffiltext{17}{Astronomical Institute, University of Utrecht, Princetonplein 5, 3584 CC Utrecht, The Netherlands}
\newpage
\begin{abstract}
As part of the Panchromatic Hubble Andromeda Treasury (PHAT) multi-cycle program, we observed a $12^\prime \times 6.5^\prime$ area of the bulge of M31 with the WFC3/UVIS filters $F275W$ and $F336W$. From these data we have assembled a sample of $\sim$4000 \uvb, old stars, vastly larger than previously available.  We use updated Padova stellar evolutionary tracks to classify these hot stars into three classes: Post-AGB stars (\pagb), Post-Early AGB (\peagb) stars and \AGBm\ stars. \pagb\ stars are the end result of the asymptotic giant branch (AGB) phase and are expected in a wide range of stellar populations, whereas \peagb\ and \AGBm\ (together referred to as the hot post-horizontal branch; \hphb) stars are the result of insufficient envelope masses to allow a full AGB phase, and are expected to be particularly prominent at high helium or $\alpha$ abundances when the mass loss on the RGB is high. Our data support previous claims that most \uvb\ sources in the bulge are likely hot (extreme) horizontal branch stars (\hhb) and their progeny. We construct the first radial profiles of these stellar populations, and show that they are highly centrally concentrated, even more so than the integrated UV or optical light. However, we find that this \uvb\ population does not dominate the total UV luminosity at any radius, as we are detecting only the progeny of the \hhb\ stars that are the likely source of the UVX. We calculate that only a few percent of MS stars in the central bulge can have gone through the \hphb\ phase and that this percentage decreases strongly with distance from the center.  We also find that the surface density of hot \uvb\ stars has the same radial variation as that of low-mass X-ray binaries.  We discuss age, metallicity, and abundance variations as possible explanations for the observed radial variation in the \uvb\ population.
\end{abstract}


\section{Introduction}
\label{intro}
Many elliptical and large spiral bulges show enhanced ultra-violet (UV) flux towards their centers. This excess from $\sim2000$\AA\ to the Lyman limit is referred to as either the UV upturn, UV-rising branch, or UV-excess (UVX). First discovered by \citet{Code69}, explanations of the UVX converged on stellar phenomena as higher resolution telescopes came online \citep{Bertola95,King92,King95,Brown98,Brown00} and far-UV spectra from HUT and FUSE became available \citep{Brown97}. 

The UVX is generally believed to be due to core helium burning stars (commonly known as horizontal branch stars; HB) and their descendants. These stars have evolved off the zero-age horizontal branch (ZAHB) and are following some combination of three evolutionary channels to becoming white dwarfs, depending on the fraction of their stellar envelope lost while they were red giants.  Stars with a modest amount of mass-loss on the red giant branch (RGB) will leave the red side of the HB, lose a large convective envelope as canonical asymptotic giant branch (AGB) stars, and become Post-AGB (\pagb) stars. Intermediate amounts of mass loss will leave stars with smaller convective envelopes (which will be subsequently lost) and hotter effective temperatures than canonical AGBs. These stars will leave the AGB track early, becoming post-early AGB (\peagb) stars\footnote{This same effect can also arise in certain scenarios where stars burn their envelopes from the bottom up on the ZAHB, also leading to \peagb s.}. Stars that lose nearly all their envelopes on the RGB become extreme (blue) horizontal branch (EHB) stars. These stars will have envelopes which are too small and temperatures which are too high to reach the canonical AGB line, and instead will become \AGBm\ or post-EHB stars \citep[for a full review see][]{Greggio99,Oconnell99}. 

\begin{figure}
\includegraphics[width=\columnwidth]{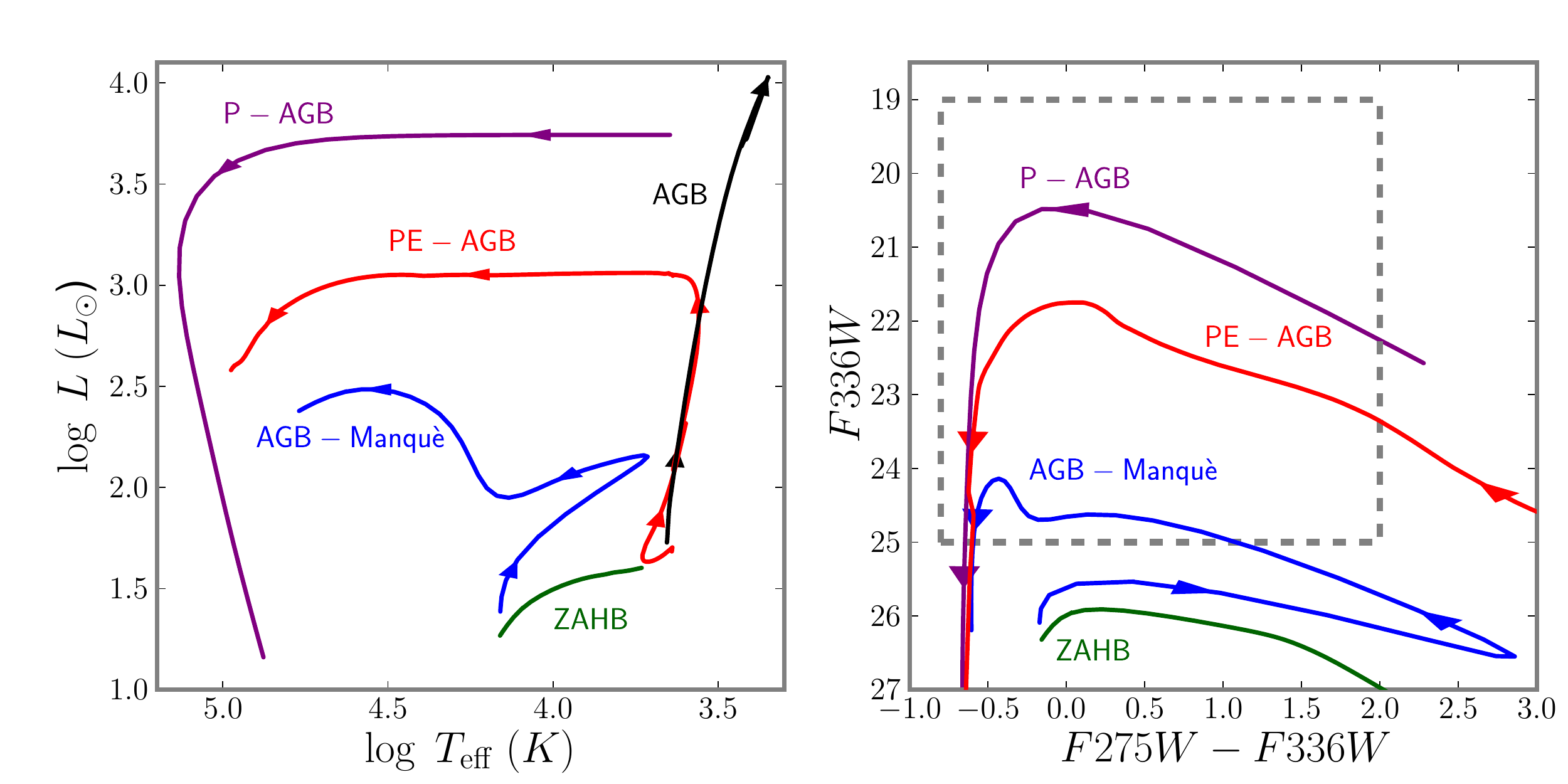}
\caption{Sample stellar evolutionary tracks \citep{Bressan12} showing hot HB stars and their progenies' placement on HR diagram (left) and CMD in UVIS filters (right) as seen at the distance of M31 (see Section \ref{sec_cmds}). Each evolutionary channel is illustrated and labeled. A canonical AGB track (black) is not shown on right panel as it is fainter than $F336W = 28$. The \pagb\ track is taken from \citet{Vassiliadis94} H-burning tracks for a mass of 0.597 \Msun\ and Z=0.016. Dashed box shows approximate range of other CMDs in this work.}
\label{f1}
\end{figure}

Examples of these stars' evolutionary tracks and the hot side of the ZAHB are illustrated in Figure \ref{f1} for both a Hertzsprung-Russell diagram and UV color-magnitude diagram (CMD). \AGBm\ and \peagb\ stars collectively are long-lived, hot post-HB stars (hereafter referred to as \hphb). These \hphb\ are 2-4 magnitudes brighter in $F336W$ than \hhb\ stars with similar $F275W-F336W$ color. \hphb\ stars have bolometric luminosities around $10^{2-3}\ \Lsun$ and have typical effective temperatures of 15,000--20,000~K, compared to $L\sim30\ \Lsun$ and $\Teff= 10^4$~K for HB stars \citep{Bressan12}. The initial masses of stars that become \hphb\ is never more than 2 \Msun, corresponding to main sequence (MS) lifetimes $\gtrsim0.6$ Gyr.

\citet{Brown98} \citepalias[hereafter][]{Brown98} obtained HST Faint Object Camera (FOC) data in the M31 bulge, providing one of the most detailed discussions of these different types of \uvb\ stars. However, these data only covered a small area ($14^{\prime\prime} \times 14^{\prime\prime}$; see magenta box in Figure~\ref{f2}). Since radial gradients in age and metallicity are common in ellipticals and the bulges of large spiral galaxies \citep[e.g.,][]{Carollo93,Davies93,Trager00,Peletier90,Gorgas90,Thomsen87,Wirth81}, we should expect radial gradients in the UV source population as well.  These gradients can be used to constrain how age and metallicity affect the stellar evolution leading to UV \hphb\ stars.

M31 is an excellent target for the study of radial trends in galaxy bulges. It hosts the most massive and metal rich bulge \citep[e.g.,][]{Ferguson93} that can be resolved into individual UV sources (as demonstrated by \citealp{Bertola95} and \citetalias{Brown98}), has a measured UVX \citep{Burstein88}, and estimated age-metallicity gradients derived from integrated line indices \citep{Saglia10}.

In this paper, we present observations and analysis of the UV component of the bulge of M31, the first data imaged in the UV by the Panchromatic Hubble Andromeda Treasury (PHAT) survey \citep{Dalcanton12}. Section \ref{sec_data} presents our observations and their comparison to those of \citetalias{Brown98}.  We investigate the radial properties of the UV sources and present a first broad comparison to new stellar models in Section \ref{sec_analysis}.  In Section \ref{sec_disc}, we discuss our findings in light of the possible causes of the UVX. Closing remarks are presented in Section \ref{sec_conc}. All magnitudes quoted throughout this paper are in the VEGAMAG system.

\section{The Data}
\label{sec_data}

\subsection{PHAT Observations, Resolved Star Photometry, Astrometry}

\begin{figure*}
\includegraphics[width=7in]{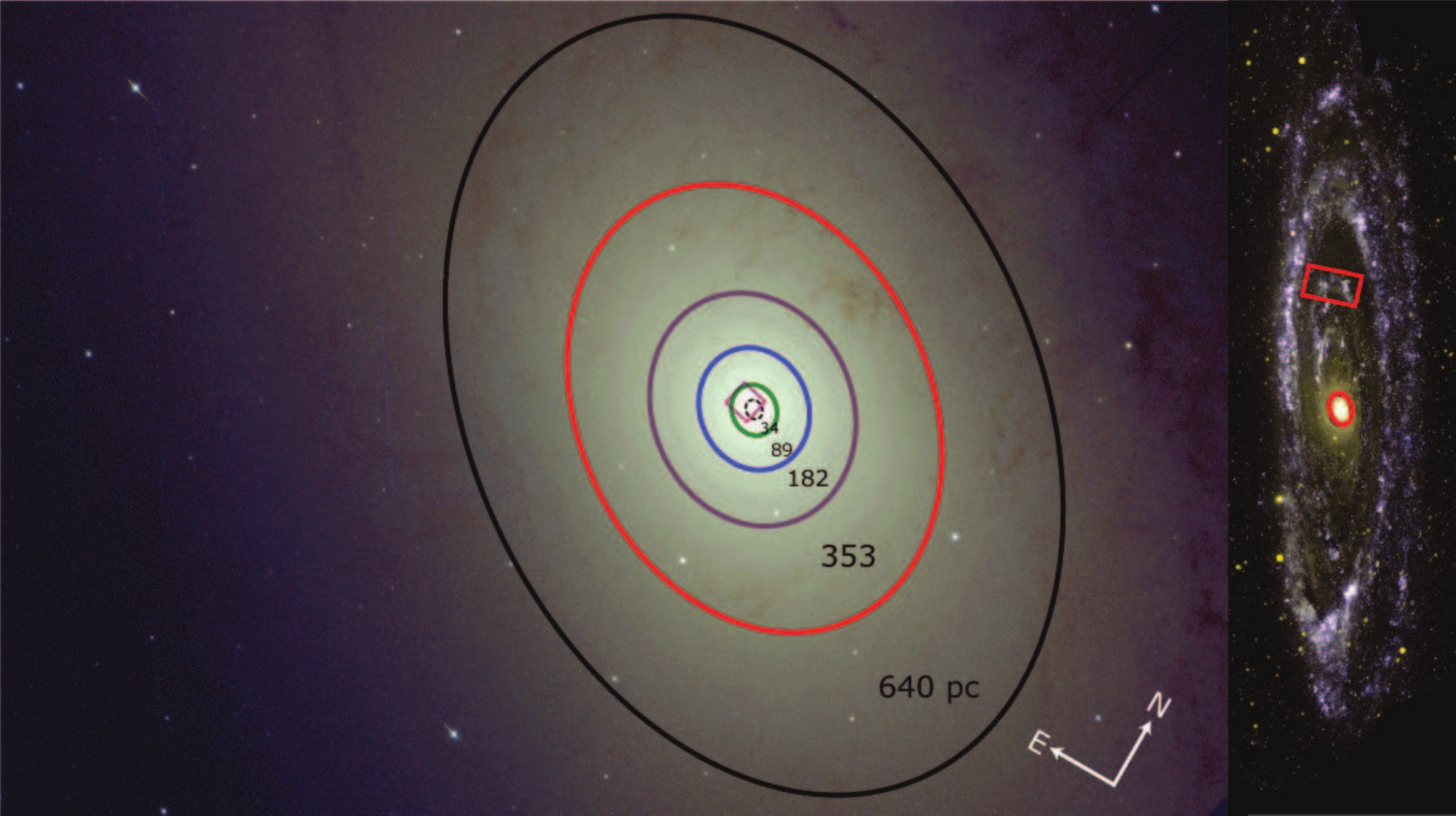}
\caption{Left: $HST$ $F336W$ (blue), $F475W$ (green), and $F814W$ (red) composite mosaic of M31 bulge.  Elliptical contours show locations of isophotes used as analysis region boundaries. The color scheme is used in other figures in this work and the median radius of each isophote bin is superimposed. Data within the innermost (black dotted) region are excluded from the analysis (see Section \ref{rad_dist}). The magenta box shows the footprint of the \citetalias{Brown98} FOC imaging. Right: GALEX FUV (blue) and NUV (green) composite mosaic of M31 \citep{Barmby06}.  The red ellipse denotes the outermost (black) contour displayed in left panel. The footprint of the M31 disk field plotted in Figure \ref{f5} is shown as a red rectangle in right panel.}
\label{f2}
\end{figure*}

\begin{deluxetable*}{crrrrrrccccc}
\setlength{\tabcolsep}{0.02in}
\tabletypesize{\scriptsize}
\tablecaption{Photometry Measurements}
\tablehead{\colhead{Region} & 
    \colhead{Region} & 
    \colhead{Region} & 
    \colhead{Region} & 
    \colhead{median} & 
    \colhead{median} & 
    \colhead{$N_{\rm stars}$} & 
    \colhead{90\%} & 
    \colhead{90\%} & 
    \colhead{Frac} & 
    \colhead{Frac} \\ 
    \colhead{} &  
    \colhead{area} &
    \colhead{$a_{\rm min}$} & 
    \colhead{$a_{\rm max}$} & 
    \colhead{$a$} &  
    \colhead{$a$} &  
    \colhead{} & 
    \colhead{Comp.} & 
    \colhead{Comp.} & 
    \colhead{Cont.} & 
    \colhead{Cont.} \\ 
    \colhead{}  & 
    \colhead{(arcsec$^{2}$)} &
    \colhead{(arcsec)} & 
    \colhead{(arcsec)} &
    \colhead{(arcsec)} &
    \colhead{(pc)} & 
    \colhead{} & 
    \colhead{$F275W$} & 
    \colhead{$F336W$} & 
    \colhead{$F275W$} & 
    \colhead{$F336W$} 
}
\startdata
0 & 76 & 0.0 & 4.3 & 2.1 & 8.2 & ... & ... & ... & ... & ... \\ 
1 & 600 & 4.3 & 13.6 & 8.9 & 34.0 & 280 & 24.3 & 23.9 & 0.27 & 0.22 \\ 
2 & 3197 & 13.6 & 33.0 & 23.3 & 88.5 & 607 & 24.4 & 24.0 & 0.12 & 0.14 \\ 
3 & 9816 & 33.0 & 63.0 & 48.0 & 182.3 & 856 & 24.5 & 24.4 & 0.04 & 0.09 \\ 
4 & 34422 & 63.0 & 123.0 & 93.0 & 353.2 & 1187 & 24.5 & 24.7 & 0.02 & 0.03 \\ 
5 & 85143 & 123.0 & 214.0 & 168.5 & 640.0 & 1404 & 24.6 & 25.1 & 0.01 & 0.01
\enddata
\tablecomments{Region area is of each logarithmically spaced isophote; $a$ is the semi-major axis (see Section~\ref{rad_dist}), given in minimum,  maximum, and median. The median radius is calculated directly from the isophotal boundary, assuming distance modulus $(m-M)_0 =  24.47$ \citep{McConnachie05}. $N_{\rm stars}$, are the numbers of UV-sources in each annulus (detected in both filters and brighter than the 90\% completeness magnitude of the innermost bin); $90\%$ Comp., percent completeness magnitude in each annulus. Frac Cont., fractional contamination at the 90\% completeness magnitude of each annulus (see Section \ref{blends}).}
\label{t1}
\end{deluxetable*}

As part of the PHAT program (GO-12058), we obtained UV imaging over a $12^\prime \times 6.5^\prime$ region ($2.6\times1.4$ kpc) around the center of M31 with HST/WFC3-UVIS. The UVIS channel has a pixel size $15$\um\ and plate scale of 0.04 arcsec/pixel. Images were taken in the $F275W$ and $F336W$ filters. Although the filters were chosen primarily for the study of massive stars in the star-forming disk, they also allow an excellent sampling of the \uvb\ sources across the bulge. The area covered for this study is illustrated in Figure~\ref{f2} and consists of a $3\times6$ grid of pointings with a $180^{\circ}$ orientation flip between the two $3\times3$ subgrids. The UVIS fields overlap by $\sim45^{\prime\prime}$. Two exposures were taken at each position in each filter, with a 1.9$^{\prime\prime}$ dither between exposures to cover the UVIS chip gap. Total exposure times are 1010 seconds and 1350 seconds in $F275W$ and $F336W$ respectively. For complete details of the PHAT observing strategy see \citet{Dalcanton12}. 

We performed point spread function photometry on all UVIS pipeline processed data ({\tt .flt} files) in the region of interest using the software package DOLPHOT\footnote{http://purcell.as.arizona.edu/dolphot}.  DOLPHOT is a modified version of HSTPhot \citep{Dolphin2000} that has been updated to include a specialized WFC3 module. Cosmic rays were rejected from the raw images using the IDL package {\tt lacosmic}, which masks cosmic rays based on their very sharp edges.  Furthermore, our sharpness cuts (see below) removed most other cosmic rays from the photometry catalog.

We have merged the independent photometric catalog for each of the 18 pointings into a single catalog for the entire bulge. We have removed duplicate sources in overlapping regions as follows. First, given DOLPHOT's source positions and initial astrometric solution, we searched the overlapping regions for pairs of stars that are close in celestial coordinates.  We used a search radius of 2 arcseconds which is large enough to compensate for HST pointing errors.  We then produced a histogram of ($\Delta$RA, $\Delta$Dec) vectors, which yielded a large but nearly uniform background of false matches plus a bump of correct matches at the RA, Dec offset between the two pointings.  We measured these offsets for all pairs of pointings, then performed a least-squares fit to find the offsets and affine corrections (scales, rotations, and shears) between the pointings.  By applying these corrections, we could put the sources detected in the separate pointings on a common local astrometric system.  We then searched for sources that are measured in multiple pointings and merged their photometric measurements so that each star was represented by a single catalog entry. The local astrometric system was then tied to a global astrometric system by first aligning the WFC3/UVIS sources with our ACS/WFC sources, which are in turn aligned to a local reference catalog produced from CFHT imaging, which is in turn tied to 2MASS \citep{Skrutskie06}.

The photometry output was then filtered to only allow objects classified as stars with DOLPHOT parameters 
{\tt signal-to-noise} $> 4$, {\tt sharp} $< 0.075$, {\tt crowd} $< 0.5$, and {\tt round} $< 1.5$ in both filters \citep[see][for a detailed description of each parameter]{Dolphin2000}. 

To measure the completeness function, we performed 100,000 artificial star tests (ASTs) for each observed field, and applied the above photometric cuts to the results. We find the data to be at least 90\% complete in each radial bin down to $F275W = 24.3$ and $F336W =  23.9$. 
We limit our analysis to this innermost 90\% completeness limit to ensure that completeness is at least 90\% at all radii. 
Completeness magnitudes are listed in Table \ref{t1} for each radial bin. We made one further cut to limit the contamination of hot Milky Way (MW) foreground stars by removing bright red stars with $F336W< 19$ and $F275W-F336W>0.8$ (see Section \ref{rad_dist}).

\subsection{Comparison with FOC data} 

As a consistency check, we compared our photometry to that of \citetalias{Brown98}, which carried out a comparable UV stellar population analysis, but over a much smaller area than presented here.  \citetalias{Brown98}'s imaging was taken by the FOC in its $F175W$ and $F275W$ filters, which are suited for the detection of stars with effective temperatures up to $\Teff \sim 40,000$ K.  To compare these datasets, we began by correcting the FOC photometry to be 0.5 mag brighter, as recommended by Section~4.4.1 in \citet{Brown00} to account for calibration revisions.  We made no conversion between the FOC and WFC3 filter systems, beyond transforming the \citetalias{Brown98} data from its tabulated STMAG values into the VEGAMAG system.  We then transformed the \citetalias{Brown98} catalog positions onto our astrometric system using linear shifts determined by matching bright sources ($F275W < 22$) in the similar $F275W$ passbands.  Once aligned, we constructed comparison catalogs with identical spatial coverage defined by the $14" \times 14"$ FOC field of view (see Figure \ref{f2}) and excluded a circular 2.5'' radius in the center to exclude the most crowded regions. Finally, we performed cross-catalog matching using a $\sim 2$ pixel ($\sim 0.07$ arcsec) search radius, and accepted matches with $F275W$ magnitude differences $< 0.7$ mag.  If multiple sources fulfilled the match criteria, we chose the match with the smallest magnitude difference.

\begin{figure}
\includegraphics[width=\columnwidth]{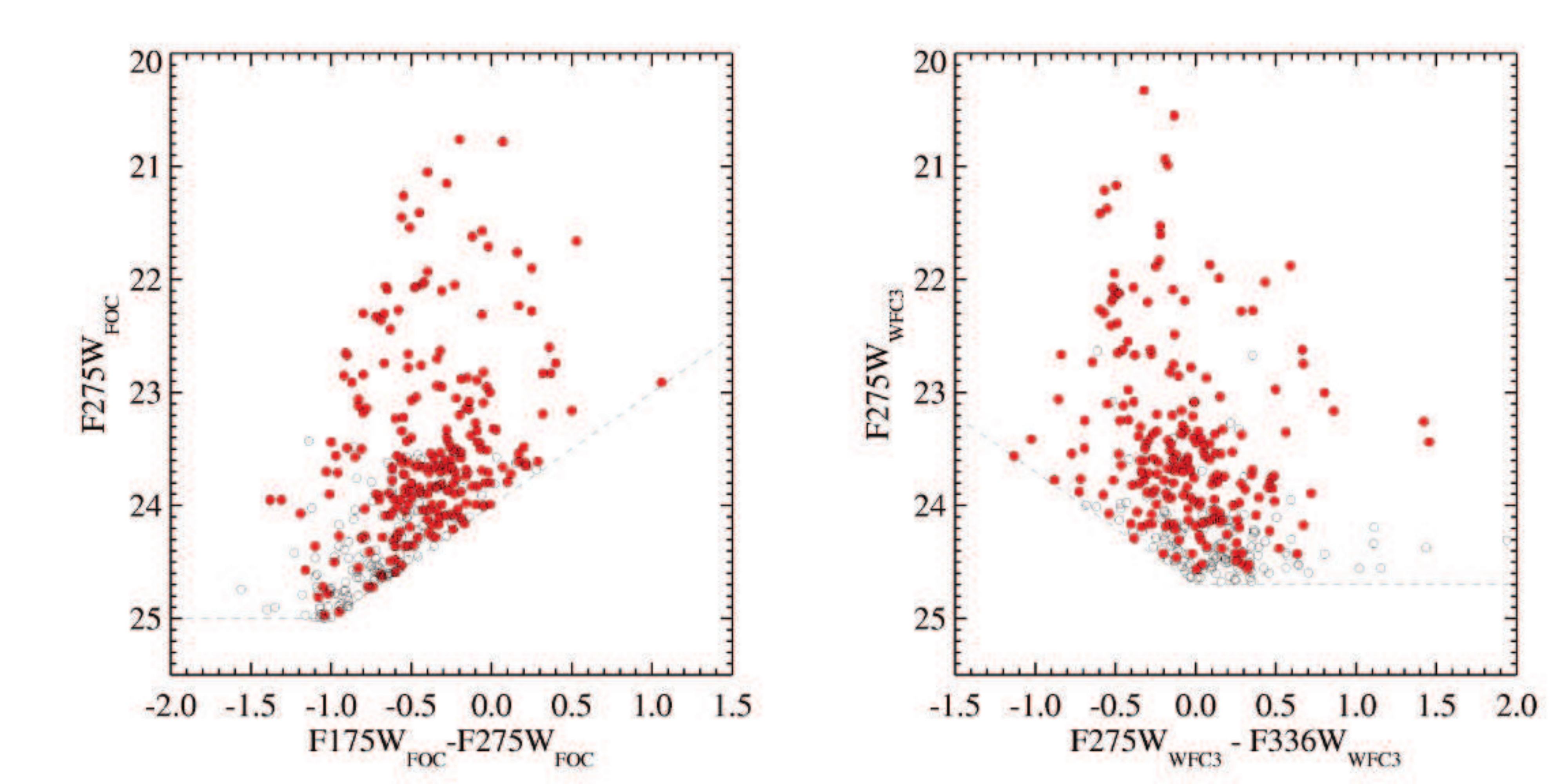}
\caption{A comparison of PHAT UV and FOC spatial overlap. Left: UV CMD for the \citetalias{Brown98}  dataset.  The FOC sources with PHAT counterparts are shown in red (filled circles), and FOC sources without PHAT counterparts are open circles.  The PHAT data appear incomplete for the bluest sources with $F175W_{FOC} - F275W_{FOC} < -0.75$.  Right: UV CMD for the PHAT dataset extracted from the smaller FOC footprint of \citetalias{Brown98}.  Stars that have matched counterparts in the \citetalias{Brown98} catalog are plotted in red (filled circles), and unmatched PHAT sources are open circles.  The completeness limit of the PHAT data is roughly 0.3 magnitudes brighter in $F275W$ than the \citetalias{Brown98} FOC observations. The dashed lines correspond to \citetalias{Brown98} detection limit.}
\label{f3}
\end{figure}

The results are presented in Figure \ref{f3}.  We were able to match 65\% of the \citetalias{Brown98} catalog to WFC3 sources in our catalog.  A majority of the unmatched \citetalias{Brown98} sources fall in the fainter ($F275W_{FOC} > 24.5$), bluer ($F175W_{FOC} - F275W_{FOC} < -0.5$) portion of the FOC CMD, showing the \citetalias{Brown98} dataset's slightly shorter wavelength sensitivity.  Overall, however, this comparison showed that these two datasets probe comparable stellar populations, with the PHAT dataset providing superior spatial coverage, and the FOC sample providing a better probe of the very hottest stars.

\subsection{Radial Binning of UV Sources \& Possible Contaminants}
\label{rad_dist}

\begin{figure*}
\includegraphics[width=7in]{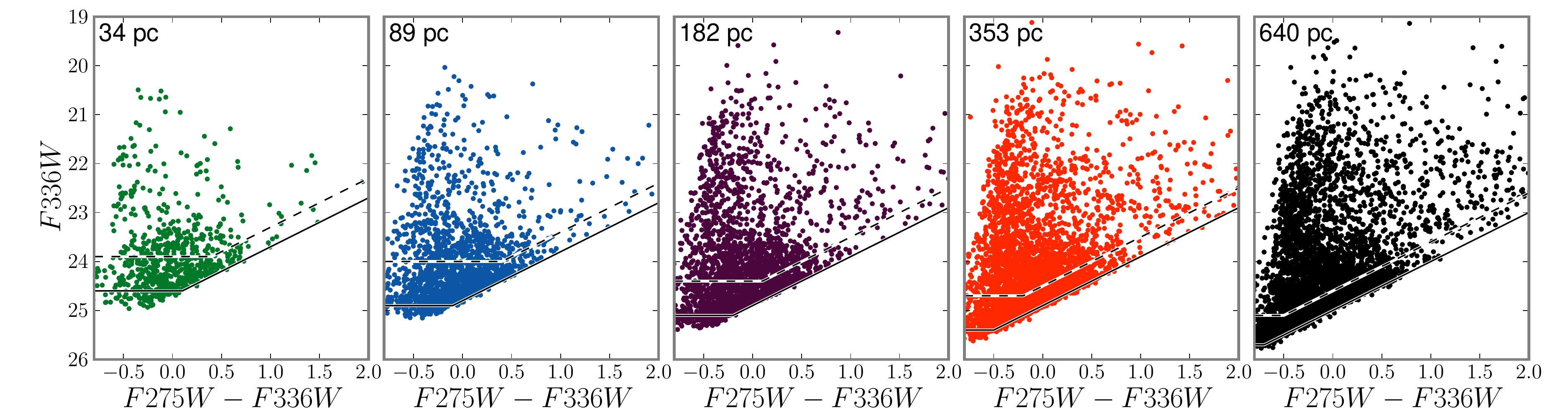}
\caption{UV Color-magnitude diagrams of each analysis region. Dashed lines mark the 90\% completeness limits and solid lines mark the 50\% completeness limits (see Section \ref{sec_data}). We limit out analysis to sources brighter than the 90\% completeness limits in the innermost (green) region. The median projected radius of each distance-corrected logarithmically-spaced isophote is labeled in the upper left. The color scheme follows Figure \ref{f2}.}
\label{f4}
\end{figure*}

\begin{figure}
\includegraphics[width=\columnwidth]{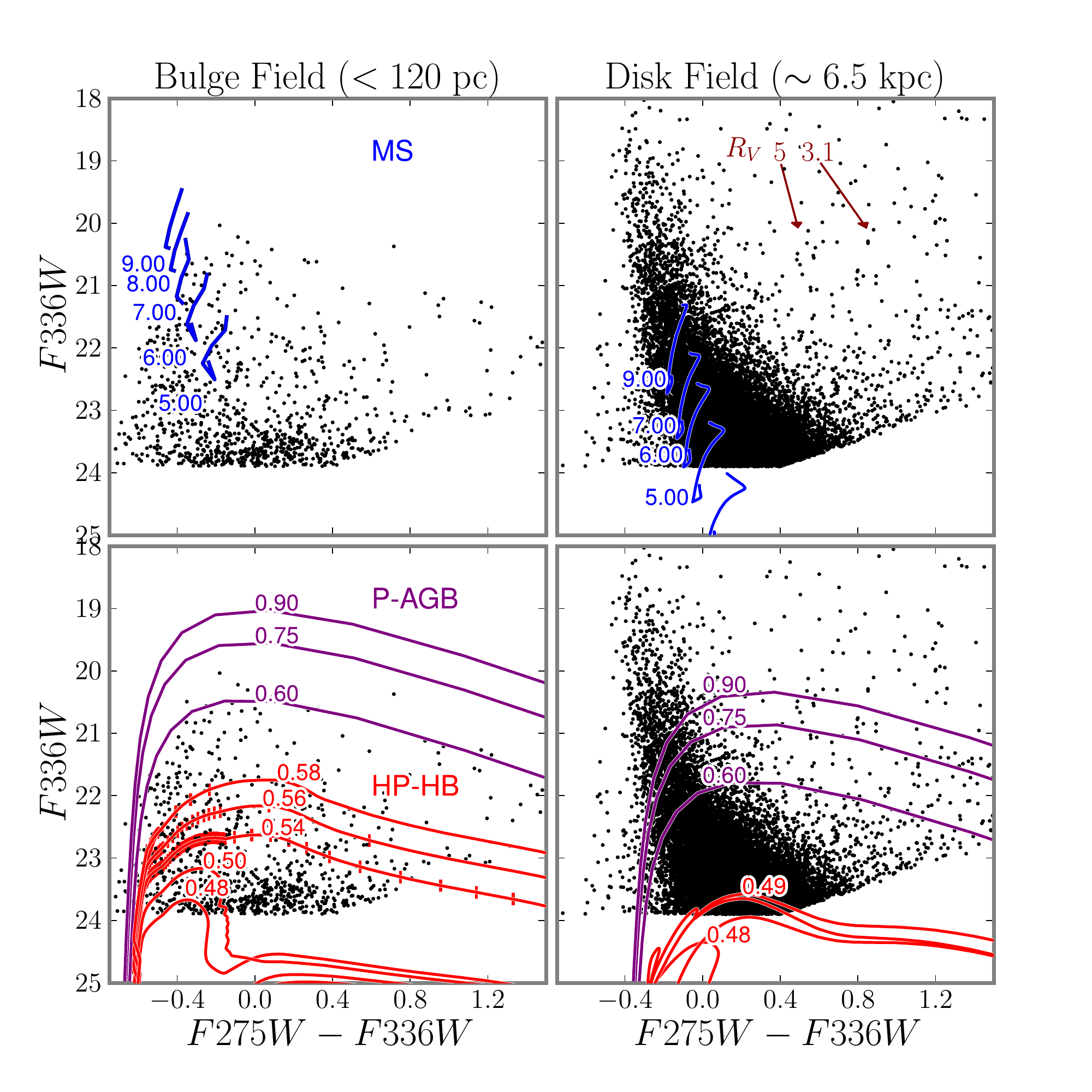}
\caption{Nature of the bright UV point sources in the central bulge (left) vs. star-forming regions in the field (right). The two panels on the left show the UV CMD for stars within 120 pc of the center; the two panels on the right show CMDs of the disk field (galactocentric radius $\sim 6.5$ kpc; see red rectangle in the right panel of Figure \ref{f2}). These CMDs are compared to MS tracks for young massive stars in the top panels; they are compared to \pagb\ and \hphb\ tracks for evolved stars in the bottom panels. It is apparent that the (reddened) MS tracks are a good match for the disk CMDs, but not the bulge (see Section \ref{sec_cmds}). Markers are placed on the three most massive \hphb\ tracks at $10^4$~yr intervals. Two reddening vectors with $R_V$ = 3.1 and 5 are also shown for 1 magnitude extinction in $F336W$. MS tracks assume $A_V =0.99$ \citep{Kang09} and bulge tracks assume a pure MW foreground extinction of $A_V =0.206$ \citep{Schlegel98}. 
\label{f5}}
\end{figure}
 
We divided the data into bins to analyze the properties of the UV sources as a function of radius.  These bins were defined using logarithmically spaced isophotes, as measured on a 3.6\um\ image from Spitzer \citep{Barmby06}.  The innermost contour (semi-major axis distance of $\sim4^{\prime\prime}$) was set to exclude the nuclear region, due to the relatively lower quality of our photometry, while the outermost contour ($214^{\prime\prime}$) was set to avoid major dust lanes and regions of recent star formation that lie along the inner arms/ring in the disk. Table \ref{t1} lists the properties of the bins, which are illustrated in Figure \ref{f2}. Their CMDs are shown in Figure \ref{f4}. 

Because metal poor MS stars can occupy the same region of CMD space as \uvb\ stars, we needed to assess potential effects of contamination on our study. We examined the contamination from the star-forming disk of M31 by comparing the radial gradients of stars in two regions of the CMD. The first CMD region was chosen to occupy the same color-magnitude space as a MS dominated disk field (galactocentric radius $\sim6.5$ kpc), shown as a red rectangle in the right hand image in Figure \ref{f2} ($22<F336W<23, -0.13<F275W-F336W<0.28$). The second CMD region was selected to be too blue to have contamination from MS stars ($22<F336W<23, -0.35<F275W-F336W<-0.50$). The stellar gradients of these regions showed a nearly constant ratio within uncertainties, indicating MS contamination from the M31 bulge is small across the analysis region. Furthermore, bulge-disk decomposition suggest that while the fraction of disk light increases five-fold between the inner and outer annuli, the disk never contributes more than 12.5\%, even in the outer most analysis region (Howley et al, in prep).

There is also a possibility that some of the redder stars, ($F275W-F336W \gtrsim 0.8$; see Figure \ref{f4}) are hot MW foreground stars. To estimate the MW contamination we have made TRILEGAL \citep{Girardi05} simulations of the MW for the field of view contained in the entire analysis region. Most of the simulated foreground stars are redder than $F275W-F336W = 0.8$ and brighter than $F336W =19$. We excluded 69 sources brighter and redder than these limits from our analysis. Bluer and fainter than these limits, our simulations suggest a negligible contribution ($\sim 40$ stars) of foreground stars with $F336W >19$.

\subsection{Integrated Light and Ancillary Imaging}
As a way of assessing bulk properties of the defined analysis regions, we calculated total integrated fluxes for each analysis region from the $F275W$ and $F336W$ imaging, as well as from the Spitzer/IRAC 3.6\um\ image.  For the HST imaging, we first assembled an aligned mosaic of the UV imaging by means of the Multidrizzle task within PyRAF \citep{Koekemoer02} using the astrometry solution obtained from the stellar catalogs and with sky-subtraction disabled. Once we obtained this combined mosaic, we performed aperture photometry of the regions on the combined image.  To estimate the sky level for the F275W and F336W mosaics, we performed photometry on outer disk images (galactocentric radius of $\sim 15$ kpc) obtained with identical observing parameters as part of the PHAT survey and confirmed that the true sky background level is $\sim0$ at these very blue wavelengths.  As such, there was no need for sky subtraction as part of our aperture photometry. For the Spitzer image, we used available mosaics \citep[see][for details]{Barmby06} and similarly performed aperture photometry for the analysis regions.

Table \ref{t2} includes the surface density of \hphb\ and \pagb\ stars (see Section \ref{sec_lumcont} for a discussion on how the populations are separated), their resolved UV flux, the total resolved UV flux, and total integrated UV flux in each filter, in each annulus. Also tabulated is the total integrated flux within each annulus obtained from Spitzer/IRAC 3.6\um\ imaging  \citep[$F_{3.6\mu\rm{m}}$,][]{Barmby06} and the bulge-to-disk ratio ($B/D$) integrated within the radial bin using a 2D, two-component model of an exponential disk and \sersic\ bulge (Howley et al., in prep).

\section{Analysis}
\label{sec_analysis}

\subsection{Radial Color--Magnitude Diagrams}
\label{sec_cmds}

Figure~\ref{f5} shows a UV CMD of the inner two radial bins (15--120 pc; left panels) and a disk field (right panels, red rectangle in Figure \ref{f2}) composed of the entire Brick 9 (galactocentric radius $\sim 6.5$ kpc) from the larger PHAT dataset. We superimposed stellar evolutionary tracks converted to the WFC3/UVIS photometric system following the \citet{Girardi08} bolometric corrections updated with the latest WFC3/UVIS filter throughputs. We assumed a distance modulus of $(m-M)_0 = 24.47$ \citep{McConnachie05}. When plotting young MS tracks, we adopted an interstellar extinction of $A_V = 0.99$ for the disk field, based on the median E(B-V) value from \citet{Kang09} for star forming regions in the disk analysis field. Although we are not currently able to derive the foreground interstellar extinction for the bulge, we adopted the value from \citet{Schlegel98}, $A_V = 0.206$. These $A_V$ values were converted into the $F275W$ and $F336W$ extinction values using the coefficients $A_{F275W}/A_V = 2.05$ and $A_{F336W}/A_V = 1.67$ derived from the \citet{Cardelli89} and \citet{ODonnell94} (shortward of $U$-band) extinction curve with $R_V=3.1$ applied to a $T_{\rm eff}=15,000$~K star.

The evolutionary tracks come from two sources: \citet{Vassiliadis94} for H-burning \pagb\ stars of masses $>0.6$ \Msun with $Z=0.016$ and $Y=0.25$ and the most recent Padova library \citep{Bressan12} for the \hphb\ stars\footnote{\pagb\ stars with masses $M>0.9$\Msun\ are not included in Figure \ref{f5} because their progenitors are too massive and short-lived, and thus not expected in the bulge.}. For the latter, the chemical composition is either $Z = 0.07$ and $Y = 0.389$ with an $\alpha$-enhanced composition typical of bulges \citep[adapted from][]{Bensby10} for the bulge field (left panels), or $Z=0.02$ and $Y = 0.285$  with solar-scaled composition for the disk field (right panels). The metallicity chosen for the bulge is an extreme case to allow for high helium content. It may be possible to populate the \hhb\ and its progeny in similar CMD space with evolutionary tracks of lower metallicity by increasing mass loss rates. The detailed analysis required to examine this possibility is forthcoming in Rosenfield et al. (in prep). The calculations for the tracks plotted in Figure \ref{f5} include {\bf all} kinds of single stars that can give rise to \uvb\ stars, namely the younger MS, the intermediate-age Post-AGBs, and the old \hhb\ and \hphb\ stars.

It is evident from the left panels of Figure \ref{f5} that the CMD of the inner 15--120 pc of the bulge shows quite different characteristics than the disk CMD in the opposing panels. The disk field shows a well populated diagonal sequence that is mostly redder than $F275W-F336W = -0.3$ and that becomes systematically redder at fainter magnitudes. In contrast, the bulge contains a significant population of stars bluer than $F275W-F336W = -0.3$ that does not appear to be an under-populated version of the disk field. 

Comparing to models, the stars in the young, star-forming disk match the solar MS evolutionary tracks, as expected. However, the higher metallicity MS tracks fail to match the bulge population -- with its wide spread in the CMD for $F336W<23$, and its remarkable concentration of stars at $F336W>23$ -- indicating that the \uvb\ stars in the bulge are not primarily MS stars. In contrast, the \hphb\ stellar evolutionary tracks provide a good match to the bulge, but fail to match the disk population. This match suggests that the \uvb\ stars in the bulge are primarily a mix of \peagb\ and fainter \AGBm\ stars (red tracks in Figure \ref{f5}, red and blue tracks, respectively, in Figure \ref{f1}). While more luminous \pagb\ stars (purple tracks in Figures \ref{f1} and \ref{f5}) with masses $>0.6 \Msun$ could potentially be detected, in practice their evolution is sufficiently fast that they make no significant contribution to the UV CMD.

If we examine the CMDs across the entire bulge (Figure \ref{f4}), there is a clear radial trend. The number of stars brighter than $F336W \simeq 23$ decreases towards the center as does the area of each isophotal region.  Dimmer stars, however, do not show this trend, and in fact, their numbers remain high at all radii. Both of these trends are probably driven by changes in the relative proportions of \pagb\ and \hphb\ stars, which in turn depends on the changes in the progenitor stars as a function of radius (though the \pagb\ do not contribute to the faint gradient). We analyze these trends more extensively below.

\subsection{Radial Luminosity Functions}
\label{sec_radlf}
\begin{figure}
\includegraphics[width=\linewidth]{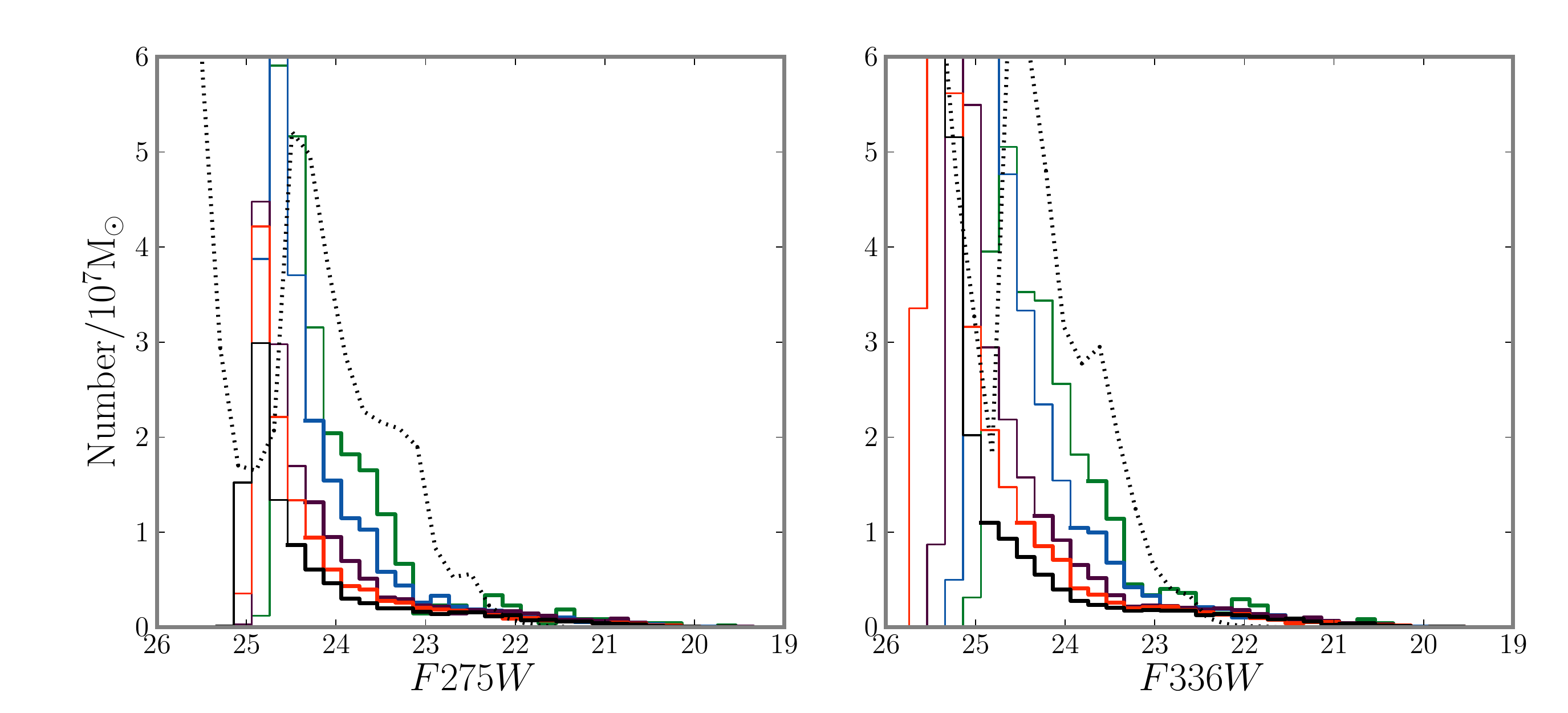}
\caption{Number of resolved \hphb\ stars (left panel: $F275W$; right panel: $F336W$) as a function of magnitude, scaled by each region's stellar mass found by the $3.6\um\ M/L$ ratio (see Section \ref{sec_radlf}), in bins of 0.2 magnitudes.  The observed LFs are corrected for completeness and shown with thin lines fainter than the 90\% completeness magnitude in the respective annulus (i.e., a correction by more than a factor of 1.1). The color scheme is the same as Figure \ref{f2}, with green indicating the innermost bin. Long lived \hphb\ stars are shown in model LF (dotted) evolved from the ZAHB with a flat distribution of mass below 1.0\Msun, and the Z=0.07 Y=0.389 tracks (cf. Figure \ref{f5}). The model is normalized to the innermost region's peak LF in $F275W$. The over-abundance of stars with $F275W>23$ in the innermost regions of M31 are associated with \hphb s. The bright LF tails are due to a seemingly flat distribution of \pagb\ stars (see Section \ref{sec_comptoother}). The increase in the model LF at faint magnitudes ($\gtrsim24.5$) is due to numerous HB stars, of which the hottest likely dominate the UVX.}
\label{f6}
\end{figure}

Figure~\ref{f6} shows the luminosity functions (LF) for contaminant-corrected (see Section \ref{rad_dist}) resolved UV sources within our adopted annuli. We removed the effects of changing radial stellar density by scaling the LF by each region's stellar mass. We estimated the stellar masses from the integrated 3.6$\mu$m flux \citep{Barmby06}, which should scale linearly with stellar mass, up to a multiplicative constant.\footnote{Isochrones from \citet[][and ref. therein]{Girardi10} confirm that the expected 3.6\um\ mass-to-light ratio does not change significantly for stellar populations of ages older than $\sim4$ Gyr and metallicities Z$ > 0.001$.} Following the procedure of \citet{Barmby06}, we assumed an optical bulge color of $B-R = 1.8$ \citep{Walterbos87}, a $K$-band $M/L$ ratio of 1.09  derived using the color-dependent relations from \citet{Bell01}, and $3.6\um$ to $K$-band $M/L$ ratio conversion from \citet{Oh08}, resulting in an $3.6\um$ $M/L$ ratio of 0.95. In Figure \ref{f6} we made completeness corrections to see if each LF peak was not simply a manifestation of incompleteness of the data. To do so, we removed our magnitude limit criteria (see Section \ref{sec_data}) and divided the sources of good quality detected in both filters by the completeness fraction found from the ASTs in 0.2 magnitude bins.  

The LFs look similar at bright magnitudes for all radii, but show significant radial variations at fainter magnitudes. First, we see a dramatic trend at faint magnitudes where faint \hphb\ stars ($F275W,\ F336W > 23$) become increasingly abundant in the inner bulge, relative to the overall stellar mass. Second, we see a weaker trend for stars with magnitudes between $21-23$, which also show increasing contributions toward the inner bulge. These are likely \pagb\ stars of a continuous distribution of masses less than $\sim0.6$\Msun, suggesting that the outer regions are under-abundant in \pagb\ stars. 

To help interpret these results, we overlay a synthetic LF (dotted in Figure \ref{f6}) made of low-mass $Z=0.07, Y=0.389$ tracks from the Padova stellar evolution library \citep{Bressan12} evolved from the ZAHB, assuming that the mass distribution of stars along the ZAHB is flat for all ZAHB masses below 1.0~\Msun. This approximation is reasonable since the range of ZAHB masses that actually contributes to the LF is very narrow ($\sim 0.45-0.58 \Msun$).

The simulated helium-burning LF reproduces the shape and color of the bump in the LF fainter than 22.5 mag, especially for the inner M31 radial bin. This bump can therefore be interpreted as being made by the \hphb s. Deeper observations would likely reveal a much larger bump in the LF due to the EHB.

In addition, the flux density from the detected stars does not vary with magnitude except for the inner region. Essentially all stars between 20 and 24 in both $F275W,F336W$ contribute equally to the flux density with the exception perhaps of region 1, where the contribution tends to increase towards fainter magnitudes.

\subsection{Luminosity Contribution of \hphb\ and \pagb}
\label{sec_lumcont}
Although \pagb\ and \hphb\ stars are individually bright in our UV bandpasses, together they only make up a small fraction of the integrated UV light ($< 2$\% in $F275W$).  A vast majority of the UV light must be emitted by fainter populations, such as \hhb\ stars that remain undetected in our resolved photometry. Theoretical calculations (Rosenfield et al., in prep.) that match the resolved star populations predict a flux contribution from faint unresolved stars that is comparable to measured flux in unresolved stars.

Since the \pagb\ and \hphb\ stars we detect are a very small fraction of the UV light (see Table \ref{t2}), the total integrated UV flux must be coming from some other population of fainter, but more numerous stars. Our model LF suggests a large population of \hhb\ stars fainter than mag $\sim 25$.  These stars are likely the main source of the UVX \citepalias[c.f.,][]{Brown98}. These stars could be detected with deeper UVIS observations and with bluer filters, provided that the crowding limit was sufficiently faint to detect these numerous stars.

To further quantify the radial gradients of stellar populations, we divided \uvb\ stars into \hphb\ and \pagb\ stars. We separated the \hphb\ stars from \pagb\ stars using the highest mass evolutionary track that produces a \hphb\ star at $Z=0.07$ and $Y=0.389$  ($M=0.58\Msun$; see the brightest red track in the bottom left panel of Figure \ref{f5}) for the Padova stellar evolution library. More massive evolutionary tracks than this critical mass produce stars with enough envelope mass to become canonical AGB stars; this behavior has little metallicity dependence, and does not appear to change with He-abundance from $0.26<Y<0.46$ \citep{Bressan12}. We have classified stars brighter than this track as \pagb.  As a faint limit of \uvb\ stars, we use our innermost region 90\% completeness magnitude (see Section \ref{sec_data}).

\begin{figure}
\includegraphics[width=\columnwidth]{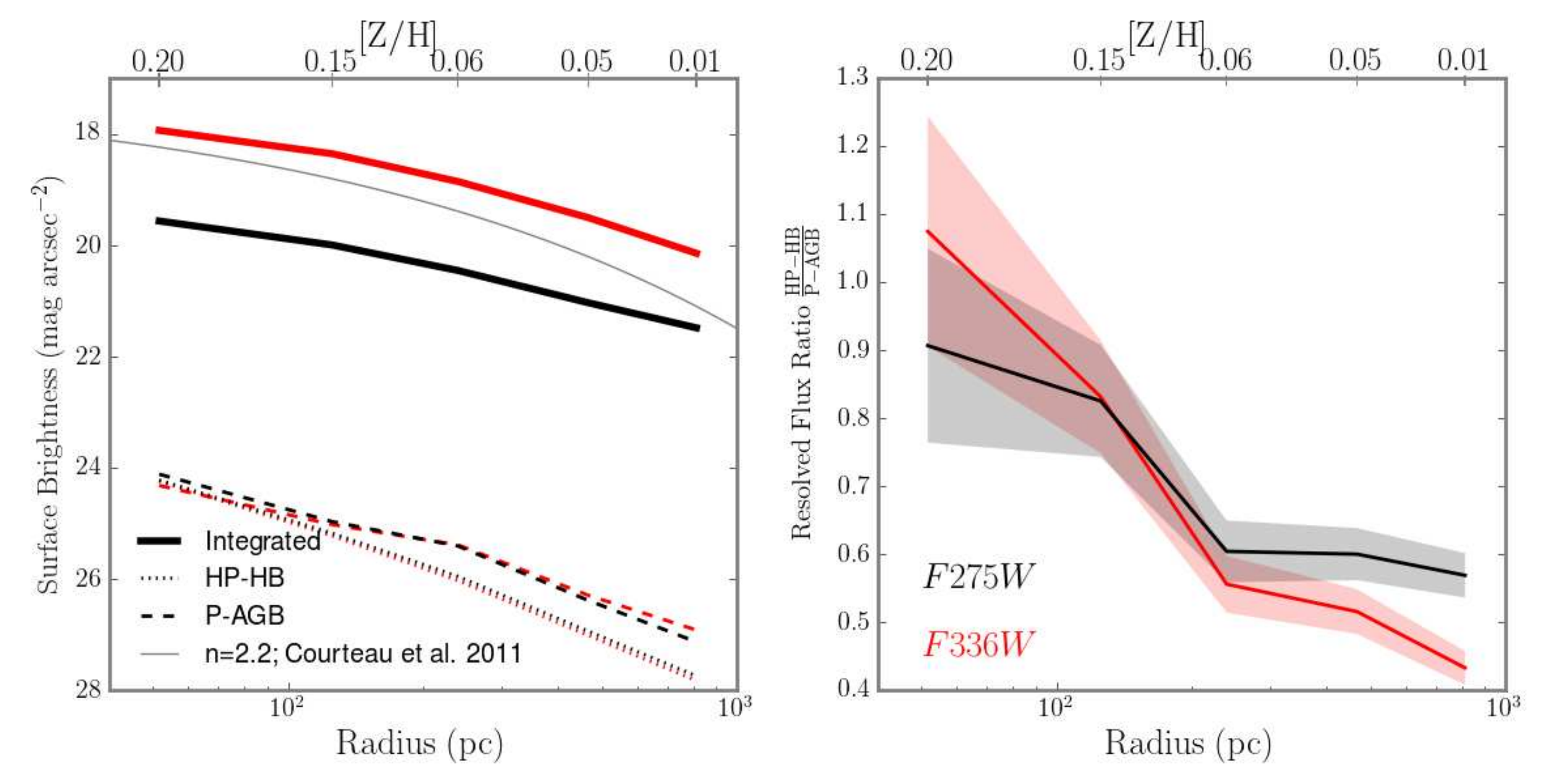}
\caption{The radial distribution of UV sources. Left Panel: Surface brightness as a function of radius of $F275W$ (solid, black), $F336W$ (solid, red), \hphb\ (dotted), and \pagb (dashed). The best fit \sersic\ profile from \citet{Courteau11} is overplotted with $\mu_e$ shifted to $F336W$=23 (gray). Right Panel: Resolved flux ratio of \hphb\ to \pagb\ stars in each filter for each annulus. Shaded regions show Poisson uncertainty in number counts.   In both panels, red denotes the $F336W$ filter and black denotes the $F275W$ filter. The characteristic metallicity as a function of radius reported by \citet{Saglia10} is indicated on the top axes. Both the \hphb\ and \pagb\ populations increase toward the nucleus of M31, but the \hphb\ stars show a dramatic increase in flux with respect to the \pagb\ stars in the innermost regions. }
\label{f8}
\end{figure}

The left panel of Figure \ref{f8} shows the surface brightness profiles of the \hphb, \pagb, and integrated UV light. In both filters, the surface brightness increases toward the center for both \hphb\ and \pagb. The right panel of Figure \ref{f8} shows the ratio of resolved UV stars, and indicates that the flux from \hphb\ stars increases with respect to the \pagb. 

A best fit \sersic\ profile from \citet{Courteau11} is overplotted on the left panel of Figure \ref{f8}, with arbitrary normalization in surface brightness. \citet{Courteau11} fit $UVBRI$, $2MASS$, and Spitzer IRAC surface brightness profiles of M31 to find the \sersic\ bulge index of $n=2.2\pm 0.3$ and effective radius of $R_e=1.0 \pm 0.2$ kpc. We did a simplistic least-squares fit to a \sersic\ profile using the surface brightness of our total integrated light in $F336W$ imagery to find $\mu_{e_{F336W}}=21.7$ mag/$\prime\prime^2$, $n_{F336W}=2.4$, and $R_{e_{F336W}}=2.1$ kpc. These numbers are consistent with the steeper scaled gradient in UV sources.

\subsection{Possible Contributions from Stellar Blends}
\label{blends}

The bulge of M31 has a strong surface brightness gradient in the UV, which must be due to a dramatic increase in the number density of stars towards the center of the galaxy. As we show in Table \ref{t2}, the majority of these stars are unresolved in our observations, but they could possibly lead to the detection of spurious sources, if sufficient numbers of the fainter stars were blended together to rise above our detection threshold. These spurious sources would dominate at fainter magnitudes and would be expected to show a strong radial gradient due to the increased crowding in the center, and thus could potentially mimic our observed gradient. While we have minimized the likelihood of this contamination by excluding the most crowded regions of the bulge and by choosing a relatively bright limiting magnitude for our analysis ($F336W=23.9$, $F275W=24.3$), it is still worthwhile to confirm that this issue is not affecting our conclusions.

\begin{figure}
\includegraphics[width=\columnwidth]{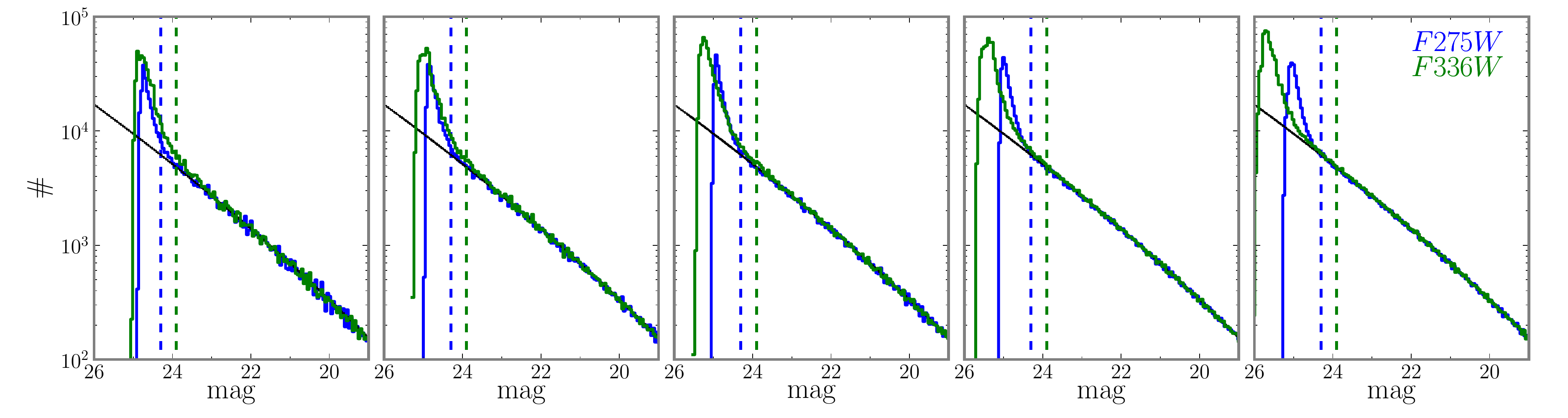}
\caption{
Input (black) and recovered (blue: $F275W$; green: $F336W$) luminosity functions in each of the radial bins. Close to the detection limit, blending causes an artificial upturn in the luminosity function.  The effect only becomes appreciable below our adopted magnitude cuts (dashed lines) and is never above 27\% (see Table \ref{t1} for exact values in each radial bin).
\label{crowdtest}}
\end{figure}

We have ruled out this possibility by considering two different cases. For the first, we assume that the luminosity function observed at bright magnitudes continues as a power-law to faint magnitudes. This assumption puts large numbers of stars just below the detection threshold, where the chances are maximized that a star will blend with an undetected source and rise above the detection threshold.  We then simulate the observed luminosity function as follows.  In a series of magnitude bins, we randomly select stars from the same catalog of artificial stars used to calculate our completeness limits.  The number of stars drawn in each magnitude bin is determined by our assumed power-law luminosity function.  We then use the recovered properties of the artificial stars (i.e., the magnitudes the stars were recovered with, and if they were detected at all) to generate the luminosity function that would be observed.  Figure \ref{crowdtest} shows the input and recovered luminosity function in each of the radial bins, plotted as black and colored lines, respectively.  As expected, blending leads there to be an artificial upturn in the luminosity function as one approaches the detection limit, with the effect being more dramatic in the most crowded inner annulus.  However, these effects only become appreciable below our adopted magnitude cuts (vertical dashed lines).  At our adopted analysis limit of $F336W=23.9$ and $F275W=24.3$, the observed luminosity function never has more than 27\% contamination from blends (see Table \ref{t1} for exact contamination levels in each bin).  Given that the luminosity function is observed to vary a factor of 100, it is unlikely that the observed radial trend is produced by radially-dependent blending from a power-law luminosity function.

We have also considered a second, somewhat less likely source of blends.  It is believed that the majority of the unresolved flux in our images comes not from sources just below our detection threshold, but from a large population of long-lived EHB stars \citepalias[e.g.,][]{Brown98} that are more than a magnitude fainter than our detection threshold.  If so, these sources must have a very high surface density, which could sometimes lead to multiple EHB stars blending together within a single resolution element.  Assuming a typical EHB magnitude of $F336W = 26.5$ \citep{Bressan12}, and a surface brightness of $\sim$18 magnitudes arcsec$^{-2}$, we can calculate the mean number of EHB stars per square arcsecond.  Adopting this mean, we can then calculate the Poisson probability that there will be a sufficiently large upward fluctuation in the local density (within the HST resolution element) to produce a spurious source at $F336W=24$. We then multiply by the number of independent resolution elements to calculate the contamination in the inner most annulus.  When we do so, we find that we expect no more than 0.1 spurious stars brightward of $F336W = 24$, in contrast to the 164 stars found between 23.5 and 24 magnitudes in region 1.  We therefore rule out blends from EHB stars as a significant source of contaminants to the observed luminosity function.

These two tests indicate that it is highly unlikely that false blended sources are significant contaminants brighter than the magnitude limit we adopted for our analysis.

\subsection{Comparison to Other Evolving Populations}
\label{sec_comptoother}
Figure \ref{f8} establishes that the resolved flux of \hphb\ increases with respect to the \pagb\ sources toward the nucleus. We now look in depth at the radial variations of these populations and expand the discussion to include other stellar populations.

The \hphb\ sources discussed here are not the first population of astrophysical sources to show enhancement towards the inner regions of M31's bulge.  In their study of Chandra X-ray point sources, \citet{Voss07} show an increase in the low-mass X-ray binary (LMXB) population at small radii within the bulge of M31.  These authors conclude that this enhancement can be accounted for by a population of dynamically produced LMXBs, which explains why the distribution's profile follows the square of the stellar density ($\rho_{*}^{2}$).

\begin{figure}
\includegraphics[width=\columnwidth]{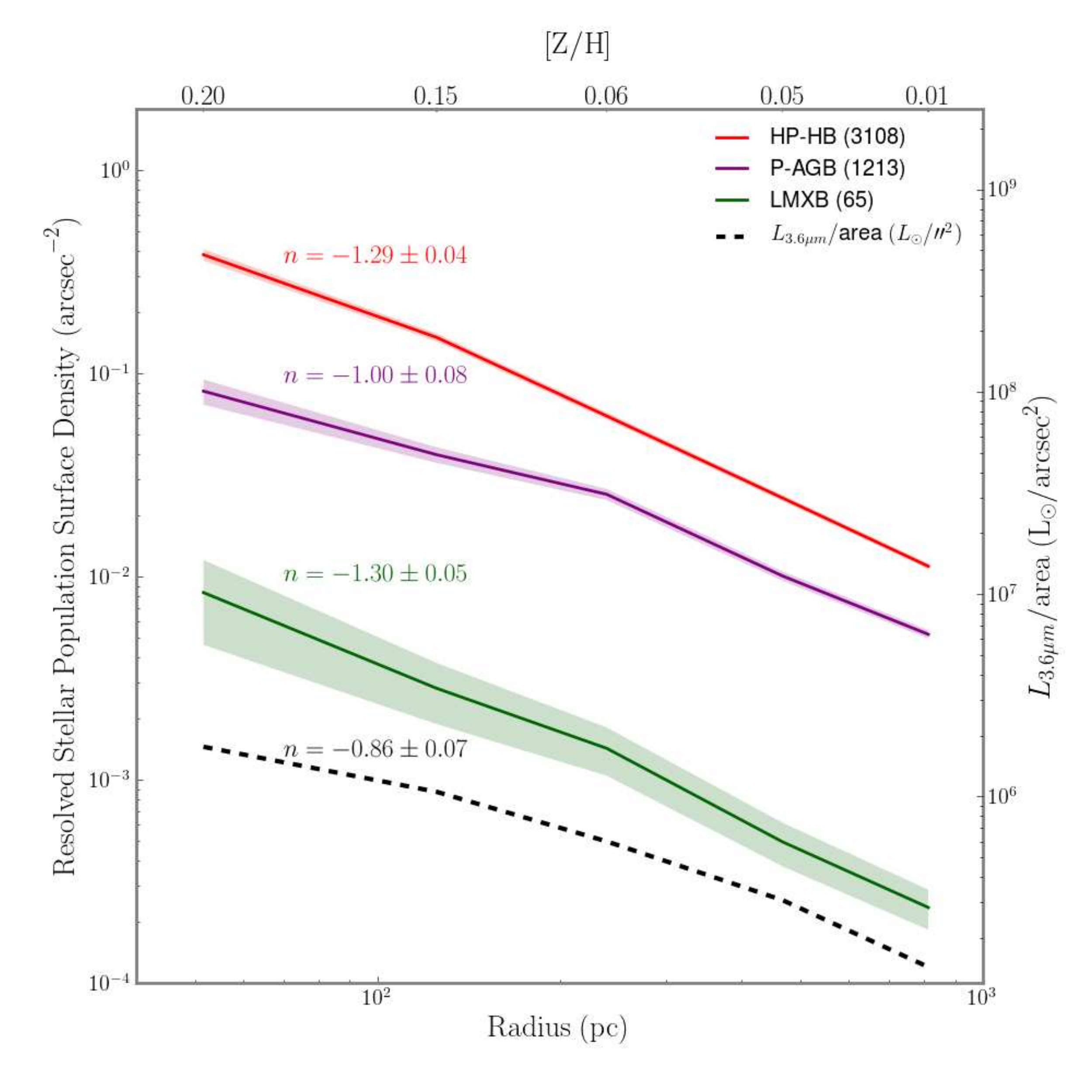}
\caption{Surface density of \uvb\ sources (blue = \hphb, red = \pagb), LMXBs \citep[green,][]{Voss07}, binned according to analysis region with Poisson uncertainties. The 3.6\um\ luminosity (black dashed) is plotted in different units and follows the right axis. The characteristic metallicity as a function of radius reported by \citet{Saglia10} is indicated on the top axis. The least-squares fit power law index of each stellar component is noted above its profile with its uncertainty in the fitting. \hphb\ (blue) and \pagb\ (red) stars follow a different surface density profile than that of the 3.6\um\ light. However, the \hphb\ stars seem to follow a profile consistent with the LMXBs.}
\label{f9}
\end{figure}

For comparison, we binned the X-ray sources from the \citet{Voss07a} catalog according to our analysis regions, excluding those sources that are either identified as non-LMXBs or that are associated with globular clusters (as their origin is different than that of unclustered LMXBs).  The resulting number density distributions for the LMXBs are shown in Figure \ref{f9}.  We find the intriguing result that the excess in \uvb\ sources is consistent with the power law index of the LMXBs.

We have also evaluated the radial trends in the density of PNe, as these stars are likely descendants from the \pagb\ populations and might be expected to show a similar radial trend. We use the \citet{Merrett06} catalog of 2615 PNe in M31 detected by 5007\AA\ emission and find that the distribution of PNe across the bulge is roughly constant with radius, and does not follow the trends of \hphb, \pagb, or LMXBs. However, the Merrett catalog is not expected to be complete in the innermost regions. If the true radial distribution of PNe does follow the gradient in the \pagb\ population, then the inner regions must be incomplete by a factor of $\sim 5.5$. We discuss the implications of these results in Section \ref{sec_disc} below.

\begin{deluxetable*}{crrrrrrrrrr} 
 \setlength{\tabcolsep}{0.02in} 
 \tabletypesize{\scriptsize} 
 \tablecaption{Radial Gradients} 
\tablehead{ 
  \colhead{Region} & 
  \colhead{$B/D$} &
  \colhead{$F_{3.6\um}$} & 
  \multicolumn{2}{c} {Integrated Flux} &  
  \multicolumn{2}{c} {Resolved Flux} &  
  \multicolumn{2}{c} {\hphb\ Resolved Flux} &  
  \multicolumn{2}{c}{Surface Density}\\  
  \colhead{} & 
  \colhead{} & 
  \colhead{} & 
  \colhead{$F275W$} & 
  \colhead{$F336W$} & 
  \colhead{$F275W$} & 
  \colhead{$F336W$} & 
  \colhead{$F275W$} & 
  \colhead{$F336W$} & 
  \colhead{\hphb} & 
  \colhead{\pagb}  \\
  \colhead{} & 
  \colhead{} & 
  \colhead{(Jy)} & 
  \multicolumn{6}{c}{} & 
  \colhead{(arcsec$^{-2}$)} & 
  \colhead{(arcsec$^{-2}$)} 
} 
\startdata 
0  & 40.7 & 0.27 & ... & ... & ... & ... & ... & ...&  ... & ...  \\ 
1 & 40.7 & 1.16 & 25.26 & 98.33 & 0.96 & 0.75 & 0.46 & 0.39 & 0.383 & 0.082 \\ 
2 & 33.5 & 3.74 & 89.77 & 351.08 & 2.23 & 1.85 & 1.01 & 0.84 & 0.150 & 0.040 \\ 
3 & 20.7 & 6.77 & 181.40 & 681.08 & 4.01 & 3.45 & 1.51 & 1.23 & 0.062 & 0.025 \\ 
4 & 12.0 & 12.99 & 373.58 & 1315.45 & 5.70 & 5.12 & 2.14 & 1.74 & 0.024 & 0.010 \\ 
5 & 7.0 & 16.71 & 608.05 & 1802.78 & 6.96 & 6.73 & 2.52 & 2.03 & 0.011 & 0.005 
\enddata 
\tablecomments{First two columns are the same as in Table \ref{t1}, followed by $B/D$, the bulge-to-disk ratio of luminosity within each annulus (Howley et al., in prep); $F_{3.6\um}$, region integrated flux within each annulus obtained from Spitzer/IRAC 3.6\um\ imaging \citep{Barmby06}. Next are integrated and resolved flux measurements in each UVIS filter, followed by the resolved flux of \hphb\ stars. All UVIS fluxes are in units of $10^{-15}$ erg cm$^{-2}$ s$^{-1}$ \AA$^{-1}$. Finally, the surface densities of the resolved \uvb\ populations are tabulated (cf. Figure \ref{f9}).}
\label{t2} 
\end{deluxetable*}

\subsection{Stellar Population Lifetimes Inferred from Observations}
Now that we have detected gradients in \hphb\ stars, it is useful to make a rough estimate of what fraction of stars are expected to become \uvb\ through this channel as a function of radius. \citet{Renzini86} provided a powerful tool to relate the number of evolved stars in a population to their lifetimes, with only a few conservative assumptions.  Using the fact that most fuel is consumed by stars fusing hydrogen or helium, and that the initial mass function is not too steep, they relate the number of stars, $N_j$; in any evolved state, $j$, to the average amount of time (in years) the star spends in that state $\langle t_j \rangle$:
\begin{equation}
\label{eq_1}
N_j = L_T B(t) \langle t_j \rangle\ ,
\end{equation}
where $L_T$ is the total bolometric luminosity of the stellar population in \Lsun\ and $B(t)$ is the rate of stars leaving the MS in units of \Lsun$^{-1}$ yr$^{-1}$. 

Equation \ref{eq_1} is strictly valid only for a stellar population of single age and metallicity that happens to contain \hphb\ stars. Any subregion of the M31 bulge, instead, is likely to contain a range of populations, with some spread in their ages and metallicities. Since just a fraction $f$ of these populations will likely produce \hphb\ stars (e.g. the oldest, most metal rich, and/or of highest helium content), equation \ref{eq_1} can be conveniently re-written as
\begin{equation}
\label{eq_2}
f = \frac{N_{\rm HP-HB}}{L_T B(t) \langle t_{\rm HP-HB} \rangle} \ ,
\end{equation}

\noindent where $f \equiv \frac{N_{\rm HP-HB,obs}}{N_{\rm HP-HB,pred}}$. This equation can be solved for $f$, allowing us to solve for the fraction of evolving stars that pass through the HP-HB channel. 

We have previously measured $N_{\rm HP-HB,obs}$, as listed in Table \ref{t2}. We also adopt a value of $B(t)=2.2\times10^{-11} \Lsun^{-1} \rm{yr}^{-1}$, derived from solar metallicity evolutionary models. This number should be accurate to 10\% for populations 10 Gyr and older \citep{Renzini86}. For the average time in the observed state, $t_{\rm HP-HB}$, we calculated the mean CMD crossing time of the relevant Padova stellar evolution tracks shown in figure \ref{f5} to find  $\langle t_{\rm HP-HB} \rangle \sim2\times10^6$ yr for the \hphb.

To calculate $L_T$, we convert the 3.6\um\ flux (corrected to exclude any disk contamination using $B/D$; see Table \ref{t2}) to bolometric luminosity. As mentioned above, 3.6\um\ flux traces the stellar population of the bulge, modulo a bolometric correction:

\begin{equation}
\label{eq_3}
L_T = L_{3.6 \rm \mu m} \times \frac{L_{\rm bol}}{L_{3.6\rm \mu m}} \ .
\end{equation}

\noindent We calculated the bolometric correction by simulating the integrated light of simple stellar popultations over a range of ages and metallicities using isochrones from \citet[][and references therein]{Girardi10}. As mentioned in passing in Section \ref{sec_radlf}, isochrones from \citet{Girardi10} confirm that the expected 3.6\um\ flux of star light does not change significantly for stellar populations of ages older than $\sim4$ Gyr and metallicities Z$ > 0.001$.

With the bolometric correction, ${\rm BC} = 2.60$ and adopting ${\rm M}_{\odot,\rm bol} = 4.77$ and ${\rm M}_{\odot,3.6\rm \mu m} = 3.24$ \citep{Oh08}, we converted $L_{3.6 \rm \mu m}$ to $L_T$,

\begin{equation}
\label{eq_3a}
\frac{L_{\rm bol}}{L_{3.6\rm \mu m}} = 10^{-0.4({\rm M}_{\odot,\rm bol} - {\rm M}_{\odot,3.6\rm \mu m}-{\rm BC})} = 2.7\ .
\end{equation}

Combining equations \ref{eq_2}, \ref{eq_3}, and substituting the values above for $B(t)$ and $\langle t_{\rm HP-HB} \rangle$, we solve for $f$,
\begin{equation}
f = 1.7 \times 10^3 \frac{N_{\rm HP-HB,obs}}{L_{3.6 \rm \mu m}} \ .
\end{equation}

\begin{figure}
\includegraphics[width=\columnwidth]{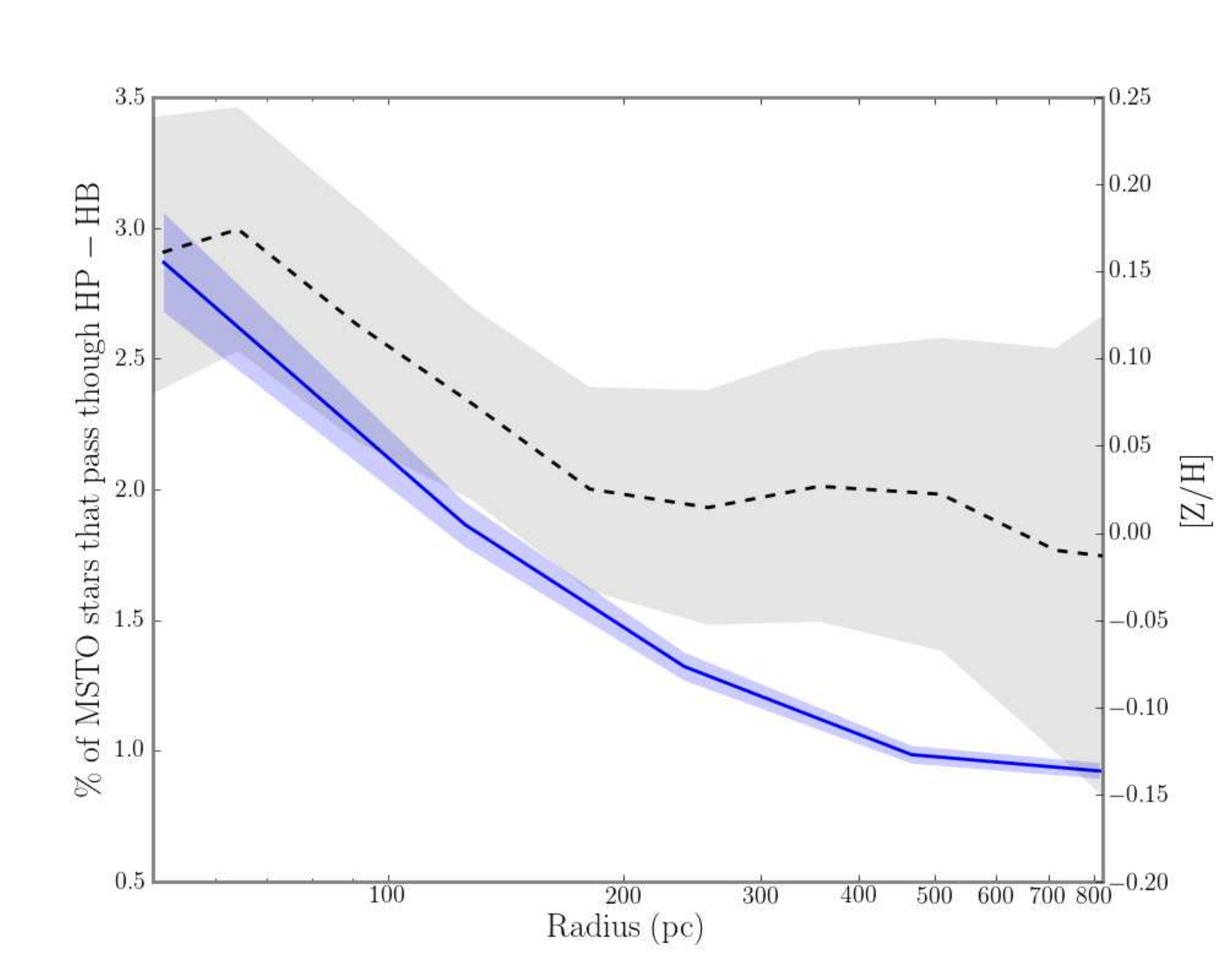}
\caption{Population variations in the central $\sim 0.5$ kpc of M31. Shown (solid, blue) is the inferred fraction of evolving MS stars that must go through the \hphb\ channel to explain the \hphb\ number count profile, given the integrated light properties. This fraction rises by about a factor of 3 towards the center. Shaded regions denote Poisson uncertainty. The Lick-Index based metallicity and its dispersion derived by \citet{Saglia10} is shown on the right axis (dotted, black).}
\label{f10}
\end{figure}

Substituting in numbers from Table \ref{t2}, we plot $f$ as a function of radius (Figure \ref{f10}). We find that the fraction of stars that go through the \hphb\ channel is small, but that the fraction increases systematically towards the center of M31.  Figure \ref{f10} shows that nearly 3\% of the MS turnoff (MSTO) stars in the innermost regions of M31 become \hphb\ stars. This implies that 97\% of the MSTO stars become canonical \pagb\ stars. In contrast, in the outermost regions of the bulge, fewer than 1\% of MSTO stars become \hphb\ stars. Note that these $\sim$3\% of HP-HB stars appearing as bright resolved UV sources belong to the same population of the fainter \hhb\ stars that, according to evolutionary models, do make a large fraction of the unresolved UV light (see Section \ref{sec_lumcont}).  

The fraction of stars that go though the \hphb\ channel in the innermost regions is consistent with what is found in M32 by \citet{Brown08}.  Our findings are also consistent with \citet{Brown97}, who demonstrated that galaxies with moderate UVX have about 2\% of the main sequence population passing through the \hhb. 

This test has shown that, even in the center of M31, very few stars from the bulge population are necessary to become \hphb\ and explain the \uvb\ stars we detect. What causes an RGB or \hhb\ star to eventually become a \hphb\ star only affects a small percentage of the underlying stellar population, but that small likelihood varies strongly with radius, for reasons we discuss below.

\section{Discussion}
\label{sec_disc}

Our LFs and CMDs allow us to confirm \citetalias{Brown98}'s claim that \hphb\ stars are the likely evolutionary channel for the production of most of the \uvb\ stars in the M31 bulge. However, $\sim20\%$ of brighter UV stars in the inner two radial bins (15--120 pc) are compatible with \pagb\ tracks (see Figure \ref{f9}), expected for stars with negligible mass loss on the RGB.

We detect clear gradients in the number and types of UV-sources. We see a steep gradient at faint magnitudes ($F275W,\ F336W \gtrsim 23$) that likely arises from an increase in the \hphb\ population towards the center of the bulge. These faint stars contribute about 60\% of the resolved flux of the inner two radial bins in both $F275W$ and $F336W$ (see Table \ref{t2}) and roughly 2\% of evolving MS stars pass though this long-lived \hphb\ channel before becoming white dwarfs (see Figure \ref{f10}), the remainder pass though the short-lived \pagb. However, the detected stars only make up $\sim$2\% of the total flux in the $F275W$ filter indicating that we have not yet detected the stars directly responsible for the UVX. These missing stars are likely to be faint \hhb\ stars, as suggested by \citetalias[e.g.,][]{Brown98}.

We now outline possible causes of the detected radial gradients of \hphb\ stars.

\subsection{Gradients in Stellar Ages}
The UVX is expected to increase with time for an aging stellar population. As a population ages, the mass of the MSTO will decrease,
so for a fixed RGB mass loss rate, the older (i.e., less massive) stars will have smaller post-RGB envelope masses and thus will be hotter HBs. As a result, the UVX has been proposed as a possible age indicator for the centers of elliptical galaxies \citep{Greggio99,Bressan94,Yi99}. \citet{Bressan94} found that the \hhb\ and \AGBm\ stars power the UV flux of a high mass galaxy at ages older than $\sim7.6\times10^9$ yr.  A negative age gradient across the M31 bulge could therefore potentially explain the gradient in \hphb\ stars we measure.

There is, however, no obvious evidence for an age gradient in the bulge of M31.  \citet{Saglia10} constrained the stellar populations in the inner bulge of M31 using spline-interpolated Lick indices models with $\alpha$-element overabundances \citep{Thomas03}, assuming simple stellar population models\citep{Maraston98,Maraston05} and a \citet{Kroupa01} IMF. They found the age of the bulge (over the regions we analyze) to be consistent with the age of the universe, with no significant radial gradient. If anything, the mean ages from \citet{Saglia10} suggest a positive age gradient with radius (i.e., older stars at larger radii), which is in the opposite sense of what is needed to produce the observed \hphb\ gradient.

Another age effect is expected if the UVX is not from \hhb\ stars and their progeny at all, but is instead from binary stars. This is an attractive explanation for why the surface density profile of LMXBs matches that of the \hphb\ stars in Figure \ref{f9}. Studies by Han and others \citep[][and references therein]{Han02,Han03,Han07} present detailed synthetic SEDs that include three binary star evolutionary channels: common envelope, Roche lobe overflow, and merging He white dwarfs. Over time, stars in these models will power the UVX.  However, if this is the case, all old populations should have a UVX, which they do not \citep[e.g.,][]{Oconnell99}. Alternatively, the lack of a central peak of PNe may suggest an alternate channel of \pagb\ evolution in the central regions, such as \AGBm\ or \peagb, which in turn produces a central peak in the \hphb\ stars. These conclusions should be considered tentative in light of the likely incompleteness of the \citet{Merrett06} PNe catalog in this region of M31.

\subsection{Galactic Metallicity Gradient}
Negative radial metallicity gradients are common features of ellipticals and large spiral bulges \citep[e.g.,][]{Roediger11,Jablonka07,Papovich01,Kobayashi99,Davies93} with typical values of $\Delta$[Fe/H]/$\Delta \log (r) \sim -0.3$ for ellipticals. Metal-rich stars are linked to more mass loss on the RGB \citep{Greggio90} which result in smaller HB envelopes and thus hotter stars and more UV flux. Thus metallicity gradients may influence the fraction of stars that enter the \hhb\ phase at each radius.

Support for the importance of RGB mass loss in producing the UVX population comes from \citet{Kalirai07}, who found that the white dwarfs of the metal-rich old open cluster NGC 6791 ([Fe/H]$\sim+0.4$, $t =  8$ Gyr) are under-massive due to the enhanced mass loss of their progenitors. The same cluster hosts a significant population of very hot helium burning stars, which, although only 30\% of all HB stars, would be sufficient to produce a UV excess. \citet{Dorman95,Buzzoni08} found with synthetic population models that only $\sim20\%$ of HB stars need to be hot to explain the UV upturn, in agreement with the NGC 6791 observations.

The small fraction of evolving MS stars that go though the \hphb\ channel could be explained by a galactic metallicity gradient if the \hphb\ are indeed the result of extreme mass loss, and the mass loss depends on metallicity. \citet{Saglia10} found that the metallicity in the bulge of M31 decreases by 0.2 dex every decade in radius, from [Z/H] $\sim 0.4$ in the center to solar at $\sim100^{\prime\prime}$ (see Figure \ref{f10}). In the inner regions of M31, it is reasonable to assume that the bulk of the stars will have metallicities that sample the metallicities reported by \citet{Saglia10}, but that there will also be a tail to high metallicities. If metallicity variations drive the gradient in \hphb\ stars, then the \hphb\ population must result from this high metallicity tail, to avoid more than $\sim$3\% of stars passing through the \hphb\ channel (i.e., Figure \ref{f10}).

\begin{figure}
\includegraphics[width=\columnwidth]{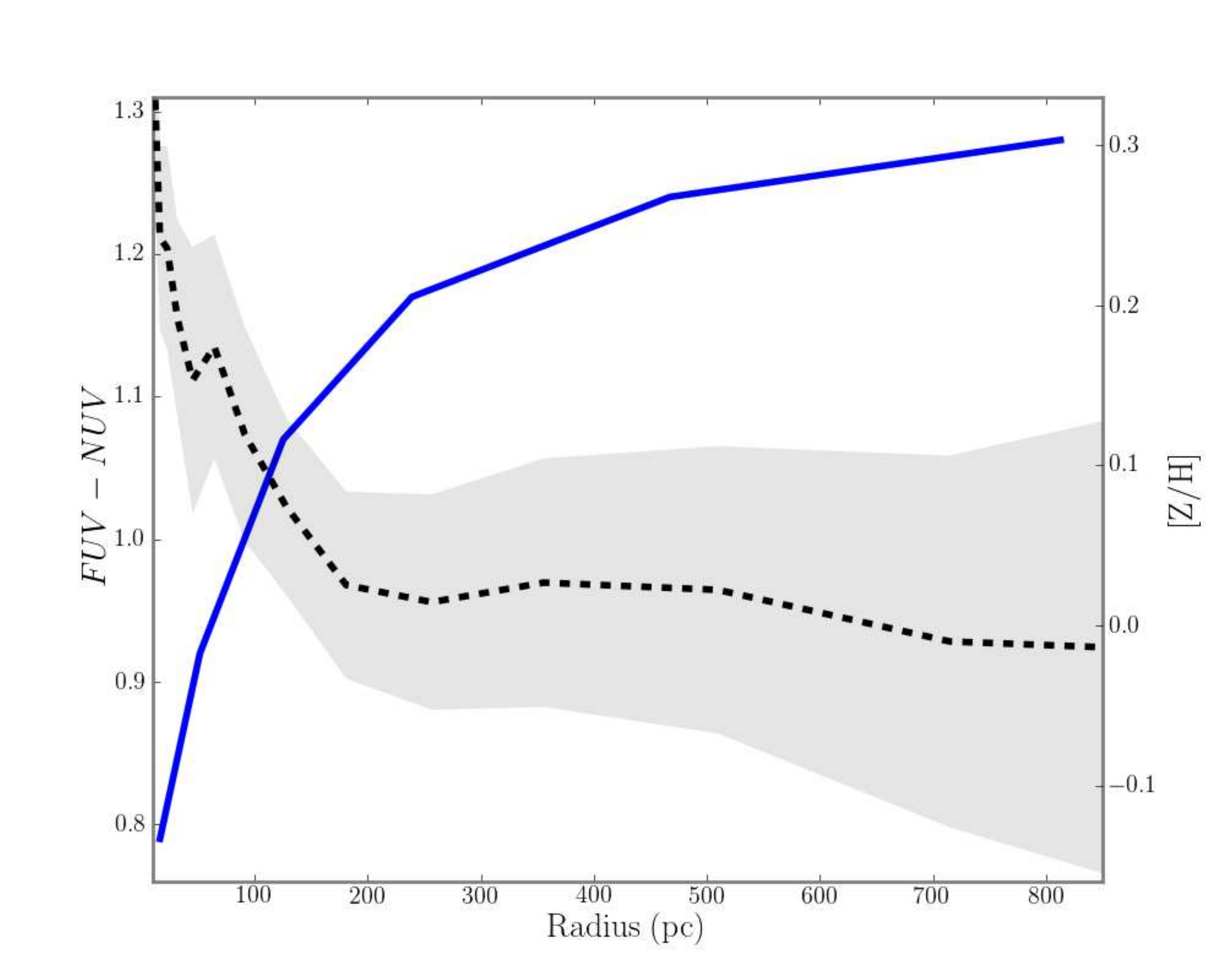}
\caption{Strong anti-correlation as a function of radius between GALEX color \citep[blue;][]{Thilker05} and the Lick Index-based metallicity and dispersion derived by \citet{Saglia10} (right axis, black). There is a sharp increase in metallicity where the $FUV-NUV$ color becomes extremely blue. This anti-correlation may be evidence of the galactic metallicity gradient driving the radial variation in the numbers of \uvb\ stars detected. The extreme blue color further shows that the \uvb\ populations are strongly confined to the inner $\sim300$ pc ($\sim75^{\prime\prime}$) of the M31 bulge.}
\label{anticor}
\end{figure}

Ancillary support for metallicity gradient being the driver of \uvb\ star production can be found in Figure \ref{anticor}, where we plot GALEX $FUV-NUV$ vs. radius, along with the metallicities derived by \citet{Saglia10}. The $FUV-NUV$ color should track the relative strength of the EHB \citep[see GALEX globular cluster CMDs in][]{Schiavon12}. It is clear that there is a sharp upturn in $[\rm{Z}/\rm{H}]$ exactly where the $FUV-NUV$ colors become extremely blue. This anti-correlation suggests that the radial variation in the numbers of \uvb\ stars follows the galactic metallicity gradient.

An alternative metallicity-dependent explanation for the UV excess is through production of low metallicity ($Z<0.008$) MSTO stars of intermediate age, which emit a large amount of their flux in the NUV. If these stars were responsible for the UV flux we detect, we would expect the UV flux to fall toward the center, due to the increased metallicity. In fact, we see the opposite gradient, the UV flux increases toward the center \citepalias[see][for a similar discussion]{Brown98}. This mechanism does not help to explain the changing fraction of \hphb\ stars.

\subsection{Elemental Abundances Gradients}
\citet{Carter11} found that the UVX depends more strongly on [$\alpha$/Fe]  than [Fe/H]. An abundance gradient in [$\alpha$/Fe] could be expected if high amounts of star formation and supernovae increase the $\alpha$-element enrichment of stars in the bulges of galaxies \citep{Greggio99,Bressan94,Catelan07}. However, \citet{Saglia10} found no obvious gradient in [$\alpha$/Fe] over the regions we analyze in M31. Within the bulge of M31, the negative radial metallicity gradient exhibits a much stronger correlation to the stellar populations responsible for the UVX. 

Higher metallicity also implies higher helium content \citetext{$\Delta Y/\Delta Z\sim2-3$, \citealt{Greggio90}; $2.5-5$, \citealt{Bressan94}; at least 2.5, \citealt{Pagel92}}. The MSTO mass decreases strongly with increasing helium abundance for a given age \citep[see][and references therein]{Oconnell99}. For a fixed amount of RGB mass loss, increased helium content will lead to hotter HB stars, that is, more \hhb\ stars. Higher initial abundances of helium will also cause the star to burn more of its hydrogen envelope during the HB phase, also favoring the production of \hphb\ stars \citep{Oconnell99,Horch92}. The increased helium abundances necessary to produce \hhb\ stars could be from contamination by the winds of older generations of AGB stars \citep{Norris04}, the winds of massive stars \citep{Decressin07}. Alternatively, these stars may have formed from the accretion of material from winds of central stars \citep{Seth10}. High helium content is a necessary ingredient to reproduce our results, and stellar winds causing the higher helium could explain the radial gradients we detect. Detailed stellar evolution modeling that varies the possible range of helium abundances from wind contamination is necessary to conclusively explain the radial gradient in MS stars that become \hphb\ stars.

\subsection{Radial Dust Gradients}
Differential dust reddening could lead to gradients in the number of \uvb\ stars detected as a function of radius. It is evident from Figure \ref{f2} that there is some structured dust obscuration in the bulge. The question then becomes, is there systematically less dust in the center of the bulge and if so, could dust alone account for the observed increase in faint UV sources towards the center of the galaxy?

Based on three lines of evidence, we find it unlikely that the UV population gradients are caused by dust.  First, M31's bulge is a 3-dimensional object, and its dust is likely to be confined primarily to the midplane. It is therefore reasonable to assume $\sim50\%$ of the bugle stars are in front of the dust layer. Second, we have inspected far-IR images of the bulge and see no systematic gradient in the FIR surface brightness over the central kiloparsec. We also see no obvious dust hole in the central regions. Finally, an $A_B$ extinction map \citep{Melchior00,Ciardullo88} shows no evidence of a significant radial dust gradient in the inner regions of M31. Therefore, while it is likely that dust has reduced the total number of UV stars we have detected (such that the percentages in Figure \ref{f10} are likely to be lower limits), it is unlikely that radial variations in extinction can be responsible for the rapid decline in \hphb\ stars with radius.

\section{Closing Remarks}
\label{sec_conc}
We have shown that the \uvb\ population in the bulge of M31 is due largely to the progeny of hot HB stars. Our findings are consistent with the conclusion of \citetalias{Brown98}, that these stars are just the tip of the iceberg of the hot HB stars responsible for the UVX. We report radial gradients in the resolved UV stellar population of M31's bulge that may be imperceptible in other data (e.g., in optical/NIR observations) because they regard less than $\sim3\%$ of the evolving stellar populations in the bulge \citepalias[$\sim2\%$ for M31's inner arcseconds and $\lesssim5\%$ for M32 cf.,][]{Brown98,Brown00}. We have demonstrated the great utility of PHAT wide-area data, though bluer, deeper data are necessary for directly measuring the UVX.

The gradients in the likely \hphb\ stars can only result from gradients in the properties of their progenitors across the bulge. The most likely changes in progenitor properties across the bulge are due to gradients in metallicity, possibly coupled to gradients in helium abundance. Detailed analysis of the most probable combinations of these parameters are in a forthcoming paper (Rosenfield et al., in prep.) where we use stellar models and population synthesis to reproduce the ratio of \pagb\ to \AGBm\ stars. This will give more precise estimates on which stellar evolution parameters are the most important in making \AGBm\ and \peagb\ stars.

\acknowledgments

This work was supported by the Space Telescope Science Institute through GO-12055.  PR wishes to express his gratitude to L.G., A.B., and Paola Marigo for invaluable guidance and for graciously hosting him in Padova. We thank Brent Groves and the M31 Herschel imaging collaboration for allowing us to examine early Herschel observations of the bulge. We are grateful to Pauline Barmby for providing us the Spitzer IRAC images. We thank Roberto Saglia for providing us with metallicity values of the M31 bulge. This research has made use of the NASA/IPAC Extragalactic Database (NED), which is operated by JPL/Caltech, under contract with NASA. The FOC data have been obtained from the Multimission Archive at the Space Telescope Science Institute (MAST). STScI is operated by the Association of Universities for Research in Astronomy, Inc., under NASA contract NAS5-26555. L.G. and A.B. acknowledge support from contract ASI-INAF I/009/10/0.

{\it Facilities:}  \facility{HST (UVIS)}.

\bibliography{phat1} 

\end{document}